\documentclass[preprint,pra,aps,showpacs,preprintnumbers,amsmath,amssymb,eqsecnum]{revtex4}

\usepackage{graphicx}% Include figure files
\usepackage{dcolumn}% Align table columns on decimal point
\usepackage{bm}% bold math
\usepackage{verbatim}
\usepackage{epsfig}
\usepackage{epstopdf}

\newcommand\beq{\begin{equation}}
\newcommand\eeq{\end{equation}}
\newcommand\bea{\begin{eqnarray}}
\newcommand\eea{\end{eqnarray}}

\def\av{{\bf a}}
\def\bv{{\bf b}}

\def\half{\frac {1} {2}}

\def\x0{{{\bf x}_0}}

\begin{document}

%\preprint{Imperial/TP/***}

%\title: NIM SchNIM}

%\title{The Leggett-Garg Inequalities and Linear Positivity: an Alternative to Non-Invasive Measurability}

\title{The Leggett-Garg Inequalities and No-Signalling in Time: A Quasi-Probability Approach}

\author{J.J.Halliwell}%
\email{j.halliwell@imperial.ac.uk}

\affiliation{Blackett Laboratory \\ Imperial College \\ London SW7 2BZ \\ UK }

%\author{Charlie Author}
% \homepage{http://www.Second.institution.edu/~Charlie.Author}
%\affiliation{
%Second institution and/or address\\
%This line break forced% with \\
%}%

%\date{\today}% It is always \today, today,
             %  but any date may be explicitly specified

\begin{abstract}
The Leggett-Garg (LG) inequalities were proposed in order to assess whether sets of pairs of sequential measurements on a single quantum system can be consistent with an underlying notion of macrorealism.
Here, the LG inequalities are explored using
a simple quasi-probability linear in the projection operators to describe the properties of the system at two times. We show that
this quasi-probability is measurable, has the same correlation function as the usual two-time measurement probability (for the bivalent variables considered here) and has the key property that the probabilities for the later time are independent of whether an earlier measurement was made, a generalization of the no-signalling in time condition of Kofler and Brukner.
We argue that this quasi-probability, appropriately measured, provides a non-invasive measure of macrorealism per se at the two time level. This measure, when combined with the LG inequalities, provides a characterization of macrorealism more detailed than that provided by the LG inequalities alone. When the quasi-probability is non-negative, the LG system has a natural parallel with the EPRB system and Fine's theorem.
A simple spin model illustrating key features of the approach is exhibited.
\end{abstract}


%02.50.Cw	Probability theory

%03.65.Ta	Foundations of quantum mechanics; measurement theory (for optical tests of quantum theory, see 42.50.Xa)

%03.65.Xp	Tunneling, traversal time, quantum Zeno dynamics

%03.65.Yz	Decoherence; open systems; quantum statistical methods (see also 03.67.Pp in quantum information; for decoherence in Bose-Einstein condensates, see 03.75.Gg)

%03.65.Ud	Entanglement and quantum nonlocality (e.g. EPR paradox, Bell's inequalities, GHZ states, etc.) (for entanglement production and manipulation, see 03.67.Bg; for entanglement measures, witnesses etc., see 03.67.Mn; for entanglement in Bose-Einstein condensates, see 03.75.Gg)

%\centerline{Imperial/TP/03-4/7}

\maketitle

\section{Introduction}

Much current research on the foundations of quantum theory has focused on the question of whether quantum theory admits a notion of realism. There are many variants on what this might mean exactly \cite{Red,Rea}, but loosely, it means that the variables describing a given situation may be regarded as possessing definite values. Together with an assumption of locality, where necessary, realism then implies that there exists an underlying probability distribution describing these variables.
Quantum theory assigns probabilities unambiguously only to sets of variables which commute. To investigate (local) realism, one can thus ask whether the set of probabilities provided by quantum theory for a set of pairs of commuting variables can be patched together into a single probability for all the variables.

The classic example of such an investigation is the EPRB set-up, which consists of a pair of particles $A$ and $B$ in an entangled state $ | \Psi \rangle $  in which their spins are highly correlated \cite{Bell,CHSH}.
Measurements are made on the spin of $A$ in directions $\av$ or $\av'$, with outcomes $s_1, s_2$ taking values $\pm 1$, and on $B$ in directions $\bv$ or $\bv'$ with outcomes $s_3,s_4$.
Quantum mechanics provides expressions for the four probabilities $p(s_1,s_3),  p(s_1,s_4), p(s_2,s_3), p(s_2,s_4)$, so, for example
\beq
p(s_1,s_3) = \langle \Psi | P_{s_1}^{\av} \otimes P_{s_3}^{\bv} | \Psi \rangle,
\label{prob}
\eeq
where the projection operators onto spin in direction $\av$ are defined in terms of the Pauli matrices by
\beq
P_s^\av = \half \left( 1 + s \av \cdot \sigma \right).
\eeq
These probabilities have the property that they are consistent amongst themselves and with the correct single spin measurement probabilities, for example
\beq
\sum_{s_1} p(s_1,s_3) = p(s_3) = \sum_{s_2} p(s_2,s_3).
\label{loc}
\eeq
Relations of this form are reflections of locality: the results of measurements on particle $B$ are unaffected by whether or not $A$ is measured. We can now ask whether the four pairwise probabilities can be regarded as the marginals of an underlying probability $p(s_1,s_2,s_3,s_4)$, so that, for example,
\beq
p(s_1,s_3) = \sum_{s_2,s_4} p(s_1,s_2,s_3,s_4).
\eeq
Clauser, Horne, Shimony and Holt (CHSH) showed that if such a probability exists then the correlation functions $C_{13}, C_{14}, C_{23} $ and $C_{24}$, defined by
\beq
C_{ij} = \sum_{s_1,s_2,s_3,s_4} s_i s_j p(s_1,s_2,s_3,s_4),
\eeq
must satisfy the eight inequalities
\beq
-2 \le C_{13} + C_{14} + C_{23} - C_{24} \le 2,
\label{CHSH}
\eeq
plus six more obtained by moving the minus sign to the three other possible locations \cite{CHSH}. The proof of this result is straightforward. Fine proved the considerably less obvious result that these eight inequalities are not just a necessary condition but also a sufficient condition to guarantee the existence of an underlying probability \cite{Fine}. (For alternative proofs see Refs.\cite{HalFine,Pit,GaMer,Bus}).
It is not hard to find quantum states for which these inequalities are violated and this has also been experimentally verified. Hence quantum theory exhibits many situations in which local realism cannot be maintained.

% Matching together patches of quasi-classicality

Leggett and Garg \cite{LG1} proposed to apply this general structure to the superficially similar but actually rather different situation of a single system described by a bivalent variable $Q$
subject to measurements at a sequence of times, $t_1 < t_2 < t_3< t_4$, described by projection operators of the form
\beq
P_s = \half \left( 1 + s \hat Q \right),
\eeq
where again $s= \pm 1$ and we have $ P_s P_{s'} = \delta_{ss'} P_s$ and $\sum_s P_s = 1$.

We again focus on set of four pairwise probabilities $p(s_1,s_2),  p(s_2,s_3), p(s_3,s_4), p(s_1,s_4)$ which are now for pairs of sequential measurements on the same system at pairs of times. (Note the different pairings of the $s_i$ compared to the EPRB case). These are usually measured in a protocol in which no more than two sequential measurements are made in each run. That is, $p(s_2,s_3)$ for example, is measured using a different set of runs to those used to measure $p(s_1,s_2)$.
%(Other protocols in which three measurements are made in the same run are %sometimes also used \cite{Jor}).

We can then ask whether there is an underlying probability for which these four probabilities are marginals. Under the rather strong assumption (about which more shortly) that these four pairwise probabilities are properly defined and compatible with each other,
the answer is again that such a probability exists if and only if the eight inequalities similar to Eq.(\ref{CHSH}) (plus six more) are satisfied, and in this context these are referred to as the Leggett-Garg (LG) inequalities,
\beq
-2 \le C_{12} + C_{23} + C_{34} - C_{14} \le 2
\eeq
(again noting the change in pairings of the $s_i$ compared to the EPRB case). Simpler versions involving just three times, analogous to the Bell inequalities \cite{Bell} are also commonly studied. A large number of papers have been written on both theoretical and experimental aspects of the LG inequalities and
a very useful and extensive recent review of the LG inequalities is that of Emary et al \cite{ELN}).

The LG inequalities were originally proposed as a test of realism at the macroscopic level, or {\it macrorealism} (MR) as it has come to be known. In practice the systems studied are rarely macroscopic, but the nomenclature is commonly used and we will follow it here. In simple terms, MR means that the system possesses trajectories in which the variables $Q$ take definite values at all four times and the four pairwise probabilities are partial snapshots of these trajectories. More precisely, the definition of macrorealism is broken down into three separate assumptions:

{\bf 1.} Macrorealism per se (MRps): the system is in one of the states available to it at each moment of time.

{\bf 2.} Non-invasive measurability (NIM): it is possible in principle to determine the state of the system without disturbing the subsequent dynamics.

{\bf 3.} Induction (Arrow of Time): future measurements cannot affect the present state.

Under these three assumptions, it has been shown in numerous places that the LG inequalities follow (see for example Refs.\cite{LG1,ELN,MaTi}). When the LG inequalities are violated it means that one of these three assumptions is false. However, what we are most interested in is violations of MRps but it is difficult to distinguish this from violations of NIM, since the sequential nature of the measurements makes NIM very hard to maintain in realistic measurements.
This assumption has been the subject of much discussion since shortly after the original Leggett-Garg paper (see for example, Refs. \cite{Ball,deco,Squid}) and a recent and very extensive critique of this and other aspects of the LG inequalities is that of Maroney and Timpson \cite{MaTi}.

%Much effort has been put into developing experimental protocols that are compatible with the NIM %assumption.

In their original work,
Leggett and Garg suggested that their inequalities could be tested in a way that respects NIM using a so-called {\it ideal negative measurement} to measure the correlation functions, in which the measuring device at the first of each pair of times couples only to, say, the $Q=+1$ state, and the absence of a detection is then interpreted to mean that the system must be in the $Q=-1$ state. As long as all measurements at the first time in each pair are measured in this way, the measurements are non-invasive since no interaction took place \cite{LG1}.

Ideal negative measurements are demanding to implement experimentally, but
there are some promising experimental results that accomplish this \cite{Knee,Rob}. An alternative experimental protocol which also claims to involve no detectable disturbance is that of George et al \cite{Geo}. There are considerable subtleties in the interpretation of these interesting results and a useful discussion of them may be found in Section 6 of the review by Emary et al \cite{ELN}.

We also note that ideal negative measurements are still invasive for a quantum-mechanical system, since, as often noted, such null measurements still involve wave function collapse \cite{Dicke} and indeed the probability for outcomes at the second time only is changed in value by this collapse, but this does not affect the value of the correlation function. What is important here is that the use of ideal negative measurements is one of a number of strategies to restrict the degree to which the measured results are explained by hidden variable theories -- it is well-established that
sets of correlation functions violating the LG inequalities can be replicated using classical stochastic models with disturbing measurements \cite{KS,Mon,Ye1,Guh}.
Or in more colloquial terms, in the background to all such experiments lurks the ``stubborn macrorealist'' who finds ingenious classical explanations of the results.
The challenge is therefore to find strategies for testing the LG inequalities which limit such classical explanations as much as possible. That is, to confirm refutations of macrorealism, we seek combinations of conditions
which can be satisfied by quantum mechanics but are very difficult to satisfy with any classical stochastic model.

An interesting recent proposal to precisely characterize the NIM requirement is the {\it no signalling in time} (NSIT) condition proposed by Kofler and Brukner, by way of analogy to the no-signalling condition in the analysis of the EPRB case \cite{KoBr}. (Similar proposals have been made earlier, for example, Refs.\cite{MaTi,Wit}). This condition reads
\beq
\sum_{s_1} p_{12} (s_1, s_2) = p_2 (s_2)
\label{cond}
\eeq
where $p_{12} (s_1, s_2) $ denotes the probability obtained under measurement at both times $t_1$ and $t_2$ and $p_2 (s_2) $ denote the probability obtained under measurement at $t_2$ only, with no earlier measurements. Corresponding conditions are assumed for the other three pairs of times.
This condition, which is regarded as a statistical version of NIM, was originally proposed as an alternative characterization of macrorealism, different to the LG inequalities and was further developed in Ref.\cite{Cle}. The NSIT condition indicates the possibility that MR can be satisfied or violated at just two times, independently of any violations at three or four times indicated by the LG inequalities.
Here we will use it in conjunction with the LG inequalities.
NSIT implies that all the two-time probabilities are compatible with each other and with the single-time probabilities. Hence if satisfied, it would have the desirable consequence that the discussion of the LG inequalities and their consequences may then proceed in a manner similar to the EPRB case.

The NSIT condition is, however, generally not satisfied by standard quantum-mechanical measurements. We briefly show why. For a system in initial state $\rho$ the probability for a single time measurement at time $t$  is
\beq
p(s) = {\rm Tr} \left( P_s (t) \rho \right)
\label{single}
\eeq
where $P_s (t)= e^{iHt} P_s e^{-iHt} $ is the projector in the Heisenberg picture. (We use units in which $\hbar = 1$).
In standard quantum measurement theory, the probability for two sequential projective measurements at times $t_1, t_2$ is
\beq
p(s_1, s_2) = {\rm Tr} \left( P_{s_2} (t_2) P_{s_1} (t_1) \rho P_{s_1} (t_1) \right)
\label{2time}
\eeq
Summing over the final measurement we have
\beq
\sum_{s_2} p(s_1, s_2) = p(s_1)
\eeq
in agreement with Eq.(\ref{single}) but summing over the initial measurement we find
\beq
\sum_{s_1} p(s_1, s_2 ) = {\rm Tr} \left( P_{s_2} (t_2)  \rho_M (t_1) \right)
\eeq
where $\rho_M (t_1) $ denotes the measured density operator,
\beq
\rho_M (t_1) = \sum_{s_1} P_{s_1} (t_1) \rho P_{s_1} (t_1)
\label{rhoM}
\eeq
which is not in general equal to the single time result, $ {\rm Tr} (P_{s_2} (t_2) \rho ) $. Hence
NSIT is not satisfied exactly in quantum mechanics except perhaps at isolated parameter values, or for specific initial states, so it is not a robust condition.

When NSIT is not satisfied exactly it is difficult to see the relationship between the LG inequalities and the existence or not of an underlying probability distribution -- if the two-time probabilities are not compatible, they cannot possibly match to any underlying probability even if the LG inequalities are satisfied.  It is then not clear what violations of the LG inequalities imply since they may come from either the failure of NSIT or the lack of an underlying realistic description, or both. For these reasons it is highly desirable to find a way to ensure that the NSIT condition, or some modification of it, is exactly satisfied.

%The approach of the present work is to first note that
The general issue at stake here is the question of finding a reasonable counterpart for sequential measurements to the probability formula Eq.(\ref{prob}) used in the EPRB case, together with reasonable conditions on it, such as NSIT, or similar. In particular, we note that when non-commuting observables are involved, there is no unique formula for the joint probabilities of such observables, or for the conditions under which these probabilities are well-defined, although this is fixed in part by the specific types of measurement contemplated. We therefore have the freedom to consider alternatives to the usual formula Eq.(\ref{2time}), perhaps subject to suitable conditions, as long as suitable measurement procedures are specified.

The main point of this paper is to show that the NSIT condition and its desirable consequences can be satisfied in quantum mechanics much more readily by switching attention from the two-time measurement probabilities Eq.(\ref{2time}) to a closely related measurable quasi-probability $q(s_1,s_2)$ with very similar characteristics as $p(s_1,s_2)$ but which, by construction, satisfies the NSIT condition Eq.(\ref{cond}) exactly.
When measured appropriately,
it describes the ``non-invaded'' aspects of the system at two times and may then be used as a non-invasive indicator of MRps.
In particular, we find that MRps holds at two times if and only if the quasi-probability is positive. When positive the desired parallel between the LG inequalities with the EPRB situation and Fine's theorem is achieved. The quasi-probability is, however, still significant when negative.
Used in conjunction with the LG inequalities, the quasi-probability permits a more elaborate characterization of macrorealism  -- it shows that MR can hold or fail at two times whereas the LG inequalities alone only give information about MR at three or four times. This leads to a more refined picture of MR (suggested already by Kofler and Brukner \cite{KoBr}).

We describe the new approach involving a quasi-probability in Section 2 and its interpretation in Section 3.
A model in which it can be successfully implemented is described in Section 4 and
we summarize in Section 5.

Finally, we mention that an elegant approach to addressing the failure of the two-time measurement probabilities to satisfy NSIT has been recently developed by Dzhafarov and Kujala \cite{DzKu}. (See also the recent discussion of this approach by Bacciagalupi \cite{Bac} and the related work by Guhne et al \cite{Guh}). In their ``contexuality by default'' approach, modified LG inequalities are derived in which the bounds on combinations of the correlation functions include terms arising from violations of NSIT.
These inequalities imply that sufficiently large violations of the LG inequalities cannot be explained by violations of NSIT and thus, the contextuality of quantum theory is cleanly distinguished from signalling or measurement effects.
A comparison of this interesting work with the present approach will be the subject of a future paper.

\section{A Quasi-probability Approach}

The two-time probability Eq.(\ref{2time}) for sequential measurements is closely linked with the quasi-probability
\beq
q(s_1, s_2) = \half {\rm Tr} \left( \left( P_{s_2} (t_2) P_{s_1} (t_1) +  P_{s_1} (t_1) P_{s_2} (t_2) \right) \rho  \right)
\label{quasi}
\eeq
In this section we discuss its mathematical properties and how it is measured.
Eq.(\ref{quasi}) is real and sums to $1$, but can be negative so is not a probability in general. (We focus on the case of the bivalent variable $Q$. Some, but not all, of what follows applies to more general variables).
The most important property of Eq.(\ref{quasi}) for what we do here is that, because it is linear in both projection operators, we have
\bea
\sum_{s_1} q(s_1, s_2) &=& {\rm Tr} \left( P_{s_2} (t_2) \rho \right) = p(s_2)
\label{con1}
\\
\sum_{s_2} q(s_1, s_2) &=& {\rm Tr} \left( P_{s_1} (t_1) \rho \right) = p(s_1)
\label{con2}
\eea
so returns the correct single time probabilities at both times, not just one.
So, unlike Eq.(\ref{2time}), the projection at the earlier time does not affect the value of the probability $p(s_2)$ at the later time. It therefore automatically satisfies a condition analogous to the NSIT condition, Eq.(\ref{cond}),
but at the expense of being negative in some regimes. We use the word ``analogous'' here since the NSIT condition refers to probabilities obtained by sequential measurements, whereas the objects used here are quasi-probabilities.
We shall shall therefore refer to Eq.(\ref{con1}) as {\it generalized no-signalling in time}.
These conditions will also satisfied by the other three quasi-probabilities of interest, $q(s_2,s_3)$, $q(s_3,s_4)$ and $q(s_1,s_4)$.

Eq.(\ref{quasi}) is one of a number of possible quasi-probabilities which match the correct marginals. It bears some resemblance to a Wigner function for finite dimensional systems \cite{Woo}, for example, but is not exactly of that form. The form Eq.(\ref{quasi}) is particularly suited to the Leggett-Garg situation, and to the measurement scheme we use, as we shall see below.
Also, we note that Eq.(\ref{quasi}) was mentioned by Marcovitch and Reznik \cite{MaRe} in their exploration of the mathematical parallels between the LG system and the EPRB situation but what we do with it here is different.

This quasi-probability is simply related to the standard quantum-mechanical two-time probability Eq.(\ref{2time}) by
\beq
q(s_1, s_2) = p (s_1, s_2)  +2 {\rm  Re} D (s_1,s_2 |-s_1,s_2)
\label{qp}
\eeq
where
\beq
D (s_1,s_2 |s_1',s_2) = {\rm Tr} \left(  P_{s_2} (t_2) P_{s_1} (t_1) \rho P_{s_1'} (t_1)  \right)
\eeq
is the decoherence functional whose off-diagonal terms are measures of interference between the two different quantum histories represented by sequential pairs of projectors.
(We use here the mathematical language of the decoherent histories approach \cite{GH1,GH2,GH3,Gri,Omn1,Hal2,Hal3,DoK,Ish,IshLin} but this is not a decoherent histories analysis of the LG inequalities).
When
\beq
{\rm  Re} D (s_1,s_2 |s_1',s_2) = 0, \ \ \ \ {\rm for} \ \ s_1 \ne s_1'
\eeq
a condition normally referred to as consistency,
there is no interference and we have $q(s_1,s_2) = p(s_1,s_2) $, and the NSIT condition Eq.(\ref{cond})
is satisfied exactly. However, noting that $p(s_1,s_2) $ is always non-negative, we see from Eq.(\ref{qp}) that $q(s_1, s_2)$ will be non-negative if the off-diagonal terms of the decoherence functional are bounded,
\beq
2 \left| {\rm  Re} D (s_1,s_2 |-s_1,s_2) \right| \le p(s_1, s_2).
\eeq
The requirement that the quasi-probability Eq.(\ref{quasi}) is non-negative
\beq
q(s_1, s_2) \ge 0
\eeq
was named {\it linear positivity} by Goldstein and Page and is one of the weakest conditions under which probabilities can be assigned to non-commuting variables, subject to agreeing with the expected formulae for commuting projectors and to matching the probabilities for projectors at a single time \cite{GoPa}.
It is satisfied very easily in numerous models, for suitably chosen ranges of parameters, since it requires only partial suppression of quantum interference, not complete destruction of it.

The quasi-probability may be expanded out as
\beq
q(s_1,s_2) = \frac {1}{4} \left(1 + \langle \hat Q(t_1) \rangle  s_1 +  \langle \hat Q(t_2) \rangle s_2  + C_{12} s_1 s_2 \right)
\label{mom}
\eeq
where
\beq
C_{12} = \half \langle \hat Q(t_1) \hat Q(t_2) + \hat Q(t_2) \hat Q(t_1) \rangle
\label{corr}
\eeq
(See Refs.\cite{HaYe,Kly} for more on this useful representation).
It will therefore be positive under the conditions
\beq
-1 + |  \langle \hat Q(t_1) \rangle  +  \langle \hat Q(t_2) \rangle |
\le C_{12} \le
1 - | \langle \hat Q(t_1) \rangle  -  \langle \hat Q(t_2) \rangle |
\label{ineq}
\eeq
By contrast the two-time measurement probability, which is always non-negative, has an extra term,
\beq
p(s_1,s_2) = \frac {1}{4} \left(1 + \langle \hat Q(t_1) \rangle  s_1 +  \langle \hat Q(t_2) \rangle s_2  + C_{12} s_1 s_2
+ \half \langle [ \hat Q(t_1), \hat Q(t_2)] \hat Q(t_1) \rangle s_2 \right).
\eeq
This extra term, which clearly vanishes when $\hat Q(t_1)$ and $ \hat Q(t_2)$ commute,
is the reason why measurements at $t_1$ affect the probability at $t_2$, since the average at $t_2$ is
\beq
\sum_{s_1, s_2} s_2 p (s_1, s_2) = \langle \hat Q(t_2) \rangle + \half \langle [\hat Q(t_1), \hat Q(t_2)] \hat Q(t_1) \rangle
\label{extra}
\eeq
This extra term is in fact the only difference between $q(s_1,s_2)$ and $p(s_1,s_2)$ and in particular note that the quasi-probability and the two-time measurement probability have the same correlation function,
\beq
C_{12} = \sum_{s_1, s_2} s_1 s_2 p(s_1, s_2 ) = \sum_{s_1, s_2} s_1 s_2 q(s_1, s_2 ).
\label{sameC}
\eeq
as previously noted \cite{Fri,MaRe} but this is not true for variables with more than two values.

There is in fact a simple physical way to understand why the correlation functions are the same. The correlation function may be written in terms of the probabilities for the two values of $Q$ being the same,
$p(same) = p(+,+) + p(-,-)$ and being different, $p(diff) = p(+,-) - p(-,+)$. We then have
\beq
C_{12} = p(same) - p(diff).
\eeq
Since the probabilities all sum to $1$ we also have
\beq
p(same) + p(diff) = 1
\eeq
Hence the correlation function is constructed from sets of histories which, although they are constructed from non-commuting operators, have zero interference, since the probabilities involved add up correctly. However, despite this essentially classical property and the similar fact that the correlation function is independent of the order of measurement, Eq.(\ref{corr}), its value can still be simulated using invasive classical measurement models so, to avoid this,
must be measured using non-invasive measurement protocols.

The two-time measurement probabilities Eq.(\ref{2time}) can be measured in a standard way. We start with the initial state $\rho$, evolve to time $t_1$, measure $Q$, then evolve to time $t_2$, measure $Q$ again. Carrying out such a run many times and noting the fraction of times the values $\pm 1$ are obtained at the two times we thus determine the probability of sequential measurement.

The quasi-probability Eq.(\ref{quasi}) can be measured most directly using sequential measurements in which the first measurement is weak \cite{weak0,weak,Wang}. For example, a weak measurement of $Q$ at time $t_1$ followed by a projective measurement at time $t_2$ will yield, for the two time probability, a term proportional to $p(s_2)$ at lowest order plus a small bias proportional to
the expression $ {\rm Re} {\rm Tr} ( P_{s_2} (t_2) \hat Q (t_1) \rho ) $. Since $ \hat Q = P_+ - P_-$, this may be written
\beq
 {\rm Re} {\rm Tr} ( P_{s_2} (t_2) \hat Q (t_1) \rho ) = q(+,s_2) - q(-,s_2)
\eeq
We also know from Eq.(\ref{con1}) that
\beq
p(s_2) = q(+,s_2) + q(-,s_2)
\label{sum}
\eeq
so if $p(s_2)$ is measured in a separate set of runs we can deduce all four components $q(\pm, s_2)$ of the quasi-probability.

However, the generalized NSIT condition Eq.(\ref{con1}) is a central condition in this approach and it would be preferable to have a protocol which actually {\it checks} this condition rather than assuming it.  This is achieved using the scheme similar to that used for ideal negative measurements described earlier.
The measuring device is first coupled weakly to only the $Q=+1$ state at time $t_1$. This weak measurement followed by a projective measurement at $t_2$ yields $q(+,s_2)$. A similar set of weak measurements are then made with a coupling to the $Q=-1$ state and we thus obtain $q(-,s_2)$. These two sets of measurements are sufficient to determine all four components of the quasi-probability. However, we then have the possibility of checking that Eq.(\ref{sum}) holds by measuring $p(s_2)$ in a separate set of runs. This is a useful check since, as noted in Ref.\cite{ELN}, weak measurements are not necessarily non-invasive.

%This procedure is of course very similar to (but not exactly the same as) the ideal negative measurement protocol commonly discussed in tests of the LG inequalities. The ideal negative measurement protocol (which, recall, is designed to eliminate certain types of classical explanations) uses, in essence, a classical argument based on a formula of the form Eq.(\ref{sum}) to deduce the probabilities in a non-invasive way.  It is sometimes noted that such an argument would not work in a quantum analysis since this formula does not hold for the two-time quantum measurement probabilities due to interference.

%Here, however, we work with the quasi-probability $q(s_1,s_2)$ which satisfies Eq.(\ref{sum}) exactly so there is no conflict between classical and quantum reasoning.

As an alternative method of measurement, we could
use the standard two-time probabilities $p(s_1,s_2 )$ to read off the correlation function and the average $\langle \hat Q(t_1) \rangle $, since these are the same for $q(s_1,s_2)$.
The average $ \langle \hat Q(t_2) \rangle $ is then measured
using a  different set of runs (i.e. not using the runs in which a measurement was made at $t_1$).
From these results $q(s_1,s_2)$ can then be constructed.
Since projective measurements are used to determine the correlation function, this method may, on the face of it, be more susceptible to alternative classical explanations (of the type outlined earlier). However, this can be avoided using ideal negative measurements to determine the correlation function.

Finally, note that although we introduced the quasi-probability in terms of the quantum-mechanical expression Eq.(\ref{quasi}), the subsequent equivalent expression Eq.(\ref{mom}) indicates that there is an alternative and more operational way of introducing it which does not involve quantum mechanics directly. This is to first measure the two averages and correlation function non-invasively, along the lines indicated above, and then to attempt to construct a probability matching them. There is of course not always a probability but one is uniquely led to the quasi-probability Eq.(\ref{mom}) which is positive in some cases. The subsequent discussion and interpretation are then the same as if we had started from Eq.(\ref{quasi}). Hence operational grounds provide an equivalent origin for the quasi-probability Eq.(\ref{quasi}).

The set of four quasi-probabilities $q(s_j,s_k)$ for $jk=12, 23, 34, 14$, subject to the generalized NSIT condition Eq.(\ref{con1}) and to Eq.(\ref{con2}), and measured according to one of the above prescriptions, are the sought-after generalization of the two-time measurement probabilities $p(s_j,s_k)$, with which we can discuss macrorealism and the LG inequalities. We therefore turn now to the interpretation of these quasi-probabilities.

\section{Interpretation of the quasi-probabilities}

We shall argue that the quasi-probabilities, properly measured, give a non-invasive measure of MRps at the two-time level, which can then be used in conjunction with the LG inequalities to characterize different types of macrorealism.

We first note that our generalized NSIT condition Eq.(\ref{con1}) and the original NSIT condition Eq.(\ref{cond}) yield essentially the same result if we average $s_2$ in both conditions, since they both indicate that
%\beq
%{\rm Tr} \left( \hat Q(t_2) \rho_M \right)=
%{\rm Tr} \left( \hat Q(t_2) \rho \right)
%\eeq
%where $\rho_M$ is the measured density operator Eq.(\ref{rhoM}).
the average $\langle \hat Q (t_2) \rangle $ is independent of an earlier measurement. This is the essential physical content of NSIT and we thus see that it is independent of whether it is expressed through a true probability or quasi-probability. The average $\langle \hat Q (t_1) \rangle $ is of course also unaffected by a later measurement as long as the induction assumption holds but this is always assumed.
Furthermore, the measurement prescriptions outlined above for the quasi-probabilities, either through weak measurements or ideal negative measurement determine the correlation function in a non-invasive way. Hence,
given the way it is defined and measured,
the quasi-probabilities may be thought of as the ``non-invaded part" of the description of the system at two times.
(Note this is not true of the usual formula, Eq.(\ref{2time}), even when measured using ideal negative measurements, since $  \langle \hat Q(t_2) \rangle $ is disturbed).

Secondly, given this description of the system at two times which satisfies NIM, the sign of the quasi-probability is then an indicator of whether MRps holds at the two time level. This is perhaps intuitively clear, but to see in more detail,
recall that the LG inequalities were derived under the key assumptions of MRps and NIM (and induction). Proceeding in exactly the same way, the same assumptions mean that we may take $Q(t_1)$ and $Q(t_2)$ to be independent random variables described by a probability. Noting that they satisfy the simple inequality
\beq
(1 + s_1 Q(t_1) ) ( 1 + s_2 Q(t_2) ) \ge 0,
\eeq
and averaging this, we obtain
\beq
1 + \langle Q(t_1) \rangle  s_1 +  \langle Q(t_2) \rangle s_2  + C_{12} s_1 s_2  \ge 0,
\eeq
which is precisely the linear positivity condition $q(s_1, s_2) \ge 0 $. Since we have argued that the quasi-probability satisfies NIM already, this shows that MRps holds at the two time level, for each of the four pairs of times, if and only if
\beq
q(s_j,s_k) \ge 0
\label{qpcon}
\eeq
for each of the four quasi-probabilities. That is, the sign of the quasi-probabilities gives a non-invasive indicator of MRps.

Consider now what this means for the various different cases. Consider first the case in which the parameters of the model are such that linear positivity Eq.(\ref{qpcon}) holds for all four pairs.
We can then ask if the four quasi-probabilities can be matched to an underlying probability $p(s_1,s_2,s_3,s_4)$, and the necessary and sufficient condition is the set of eight LG inequalities. This therefore yields a close parallel with the EPRB case and Fine's theorem. Violations of the LG inequalities in this case are then refutations of MRps at the three/four time level.

Clemente and Kofler \cite{Cle} have shown that the LG inequalities cannot, in general, provide a sufficient condition for macrorealism (although they are clearly necessary). However, the above result is not in conflict since it is not as general -- it involves a quasi-probability with a restricted set of parameter ranges.
(Fine's theorem in the LG inequalities is also discussed in Ref.\cite{MaMa}).

Although NIM is explicitly incorporated in this approach,
it is of interest to see how hidden variable explanations are restricted in their power to explain LG violations, in this case where linear positivity is satisfied.
The hidden variable models that replicate the correlation functions violating the LG inequalities have the feature that they replicate the standard two-time measurement probabilities Eq.(\ref{2time}). In particular, they are invasive and will not in general satisfy NSIT because they disturb $p(s_2)$ \cite{KS,Mon,Ye1,Guh}.
Hence the role played by (generalized) NSIT is to eliminate this type of hidden variable explanation. It is not clear, however, that it eliminates all hidden variable explanations.
We find, for example, for the case of a maximally mixed initial state discussed in the next section, that $ \langle \hat Q (t_2) \rangle $ is zero whether or not an earlier measurement is made, so $p(s_2)$ is undisturbed by the presence of an earlier measurement.
For this reason it is desirable to ensure that the correlation function is measured non-invasively in the measurement protocol. Hence non-invasiveness is ensured by a combination of NSIT and the measurement protocol.

When NSIT is satisfied and the two-time quasi-probabilities are positive,
hidden variable models of the non-invasive type may be found to describe each of the two-time situations,
but they cannot be patched together into a hidden variable model for the values of $Q$ at all four times unless the LG inequalities are satisfied. Hence hidden variable models cannot in general replicate both the quasi-probabilities satisfying linear positivity and LG inequality violation
(unless there is some very implausible collusion going on between the measurements at different pairs of times in which new invasiveness effects not present at two times appear at three or more times, as discussed, for example, in Ref.\cite{deco}). As we shall see, quantum mechanics can meet both constraints.

Suppose now that linear positivity Eq.(\ref{qpcon}) is violated, which means that at least one of the quasi-probabilities is negative for some parameter values of interest. This means that MRps has failed at the two-time level. This case has no parallel in the EPRB situation.
Since the LG inequalities are independent conditions, they may or may not be satisfied in this case. (This is not in conflict with Fine's theorem). Perhaps surprisingly, we can therefore have a situation in which MRps is violated at the two-time level, but satisfied at the three or four time level, if the LG inequalities are satisfied. This simply indicates the fact that the LG inequalities alone are not enough to fully characterize macrorealism, since there are cases such as this, where they miss MRps violations, as noted by Kofler and Brukner \cite{KoBr}.
If both linear positivity and the LG inequalities are violated then it means that MRps fails at both the two-time level and at three/four times.

In the cases where linear positivity is violated, so MRps fails at two times, it is again of interest to ask whether this may have arisen from an underlying classical model. Generally speaking, quasi-probabilities with regions of negativity often arise in situations where the system is in fact described by an underlying positive probability (i.e. satisfies MRps in our langauge) but has been rendered negative by invasive measurements. Here, however, we are not looking at general quasi-probabilities but at the very restricted class of quasi-probabilities Eq.(\ref{quasi}) which satisfy the conditions Eqs.(\ref{con1}), (\ref{con2}), one of which is our generalized NSIT condition, which specifically limits measurement disturbances and also the correlation function is measured non-invasively. So linear positivity violation cannot be simulated by classical models with disturbing measurements.

In brief, the quasi-probability Eq.(\ref{quasi}), properly measured, supplies a non-invasive measure of MRps at two times which can be used in conjunction with the LG inequalities to characterize macrorealism in number of different ways. It shows that MRps can be violated, or not, at two times, or at three or four times times. Of these perhaps the most interesting case is that in which MRps holds at the two time level but is violated at three or more times, since it is a parallel with the EPRB case and in particular, like that case, it involves on the one hand, essentially classical behaviour at two times, but on the other, involves
subtle quantum correlations which do not appear until three or more times are considered. A model exhibiting this is presented in the next section.

%We comment that Williams and Jordan argued that LG inequality violations are in %one-to-one correspondence with strange weak values \cite{Jor}. This is in apparent %contradiction with the present work since situations with $q(s_1,s_2) <0$ are %strange weak values in the weak measurement protocol described here, yet
%we concentrate on the opposite case of positive quasi-probabilities. The point %here is that Williams and Jordan worked with a protocol in which three %measurements are made in each run, some of them weakly, whereas here we only %consider two-time quasi-probabilities.

Finally, as a tangential issue, we note the following. In looking for a probability for all four variables that matches the four quasi-probabilities in the regime where they are non-negative, it would be natural to consider the four-time quasi-probability,
\beq
q(s_1, s_2,s_3,s_4) = {\rm Re} {\rm Tr} \left( P_{s_4} (t_4) P_{s_3} (t_3) P_{s_2} (t_2) P_{s_1} (t_1)  \rho  \right),
\label{quasi4}
\eeq
a possible generalization of Eq.(\ref{quasi}) to four-times \cite{GoPa}. It clearly matches the two-time quasi-probabilities so when non-negative it solves the matching problem. However, it does not have a straightforward relationship to the LG inequalities. If Eq.(\ref{quasi4}) is non-negative then the LG inequalities must be satisfied. However, if the LG inequalities are satisfied, it does not imply that Eq.(\ref{quasi4}) is non-negative.
Fine's theorem shows that if the LG inequalities are satisfied then there exists {\it some} probability matching the given two-time marginals (generally a family of probabilities), but it is not necessarily Eq.(\ref{quasi4}).
In fact an explicit example involving the EPRB set-up show that there are situations in which the CHSH inequalities are satisfied, so a probability exists, but the counterpart to Eq.(\ref{quasi4}) is negative somewhere \cite{HaYe}. Hence linear positivity for Eq.(\ref{quasi4}) is not the most general solution to the matching problem. This is not directly relevant to the main thrust of this paper and will be explored further elsewhere.

\section{A Simple Spin Model}

We now address the question of finding situations in which the LG inequalities are violated but linear positivity is satisfied for the four two-time quasi-probabilities. Mathematically, it is not hard to see how this can be achieved. For a given set of correlation functions $C_{12}, C_{23}, C_{34}, C_{14}$ violating the LG inequality four quasi-probabilities satisfying linear positivity are easily found by finding values of the averages $\langle \hat Q(t) \rangle$ at the four times satisfying Eq.(\ref{ineq}). However, the averages are not freely chosen and the issue is to find a specific model supplying the averages and correlation functions in which there are parameter ranges doing the job.

We consider a simple model involving spins which is often studied in this context (see for example Ref.\cite{Ye1}). We take the bivalent variable $\hat Q$ to be the Pauli matrix $\sigma_z$,
the Hamiltonian to be
\beq
H = \half \omega \sigma_x
\label{H}
\eeq
and the initial state to be the $| + \rangle $ state in the $z$-direction.
It is readily shown that
\beq
\langle \hat Q (t) \rangle = \cos \omega t
\label{av}
\eeq
and the correlation function Eq.(\ref{corr}) is
\beq
C_{12} =\cos \omega (t_2 - t_1)
\label{corr12}
\eeq
The LG inequalities then read,
\beq
-2 \le
\cos \omega (t_2 - t_1)
+\cos \omega (t_3 - t_2)
+\cos \omega (t_4 - t_3)
-\cos \omega (t_4 - t_1)
\le 2
\eeq
(plus six more of this form). For simplicity we take $t_1 = t, t_2 = 2t, t_3 = 3t, t_4 = 4t $ and the inequalities then read
\beq
-2 \le 3 \cos \omega t - \cos 3 \omega t \le 2
\eeq
which is maximally violated (exceeds $2$ by a factor of $\sqrt{2}$) at $\omega t= \pi/4 $. However, we must also check for linear positivity. One of the quasi-probabilities is
\beq
q (s_1, s_2) = \frac {1} {4} \left( 1+ s_1 \cos \omega t_1 + s_2 \cos \omega t_2 + s_1 s_2 \cos \omega (t_2 - t_1) \right)
\eeq
With the above choices of times, we find
\beq
q (\pm, -) = \frac{1}{4} \left( 1 - \cos 2 \omega t  \right)
\eeq
which is non-negative. However,
\beq
q(\pm,+) = \frac{1} {4} \left(  1 \pm 2 \cos \omega t + \cos 2 \omega t \right)
\eeq
from which it is easily seen that $q(+,+)$ and $q(-,+)$ always have opposite signs so one of them is always negative for all $t$.

Different choices of time intervals other than the evenly spaced one considered do not improve the situation. At some length it may be shown that, for arbitrary times $t_1, t_2$,
\beq
q(+,+) q(-,-) = \frac {1} {4} \sin \omega t_1 \sin \omega t_2 \cos^2 \left( \frac{\omega (t_1-t_2)} {2} \right)
\eeq
and
\beq
q(+,-) q(-,+) = - \frac {1} {4} \sin \omega t_1 \sin \omega t_2 \sin^2 \left( \frac{\omega (t_1-t_2)} {2} \right).
\eeq
These two expressions therefore always have opposite signs which means that there is no regime in which all four of the $q(s_1,s_2)$ are non-negative.
The above results show that it is quite easy to find situations where the LG inequalities are either violated or satisfied but linear positivity is not satisfied.

The situation with regard to linear positivity is significantly improved by taking a mixed initial state and we choose a family of such states of the form
\beq
\rho = \half (1 + \alpha \sigma_z )
\eeq
where $ | \alpha | \le 1 $.
The correlation function Eq.(\ref{corr12}) is the same (since it is in fact independent of the initial state) and the average now is
\beq
\langle \hat Q(t) \rangle = \alpha  \cos \omega t
\eeq
From this we immediately see from Eq.(\ref{mom}) that in the case $\alpha = 0$, the maximally mixed state, all the averages are zero and therefore all the quasi-probabilities are trivially non-negative since all the correlation functions satisfy restrictions of the form $|C_{12}| \le 1$. So linear positivity is always satisfied in this case.

The original NSIT condition Eq.(\ref{cond}) with the probabilities taken to be the usual quantum-mechanical ones is also satisfied exactly in this case \cite{KoBr}.
This feature has sometimes been taken to mean that there is no measurement disturbance in the case of a maximally mixed state \cite{SOS} but this is not necessarily the case \cite{KGBB}.
The fact that the averages are all zero in this case actually means that the (original or generalized) NSIT condition loses its usefulness in terms limiting hidden variable explanations -- the correlation functions could still be replicated by an invasive classical model but the tell-tale disturbances in $p(s_2)$ could be averaged to zero by the mixed initial state. Hence in this case a non-invasive measurement of the correlation function is necessary. This illustrates the statement made in Section 3 that NSIT is only a partial indicator of non-invasiveness and must be used in conjunction with the measurement protocol.
(The contextuality by default approach also has analogous features for a maximally mixed state \cite{Bac}).

Consider now more general values of $\alpha$.
The condition Eq.(\ref{ineq}) for the non-negativity of $q(s_1,s_2)$ is conveniently rewritten
\beq
\frac{ 1 \pm C_{12} } { | \langle \hat Q(t_1) \rangle \pm \langle \hat Q(t_2) \rangle | }
\ge 1
\eeq
where, to be clear the $\pm$ on top and bottom are correlated (i.e. they are both plus or both minus).
Inserting the explicit values this condition reads
\beq
| \alpha | \le  \frac { 1 \pm \cos \omega (t_2 - t_1 ) } { | \cos \omega t_1 \pm \cos \omega t_2 | }.
\label{12con}
\eeq
It is clearly always satisfied for $\alpha = 0$, as noted, but can never be satisfied for the pure state case $|\alpha| = 1$.
It can be satisfied by other values of $\alpha$ but this can impose a restriction on the ranges of the possible values of the times.

To explore this further, we choose the equally spaced time intervals described above.
%and investigate whether linear positivity can be satisfied in the neighbourhood of %the parameter values which give maximal violations of the LG inequalities.
We find that the four sets of linear positivity conditions Eq.(\ref{qpcon}), for the cases $ij =12, 23, 34, 14$ respectively, read
\bea
| \alpha | & \le &  \frac { 1 \pm \cos \omega t } { | \cos \omega t \pm \cos 2 \omega  t | },
\label{a1}
\\
| \alpha | & \le &  \frac { 1 \pm \cos \omega t } { | \cos 2 \omega t  \pm \cos 3 \omega t | },
\label{a2}
\\
|\alpha | & \le &  \frac { 1 \pm \cos \omega t } { | \cos 3 \omega t \pm \cos 4 \omega  t | },
\label{a3}
\\
|\alpha | & \le &  \frac { 1 \pm \cos \omega t } { | \cos \omega t \pm \cos 4\omega  t | }.
\label{a4}
\eea
We seek the minimum value of the right-hand side in these four inequalities. It is easily seen (e.g. by simply plotting them) that the $-$ cases have their lowest minima at $t=0$ and the $+$ cases have their lowest minima at $\omega t = \pi$.
The least of all the minima occurs for the $-$ case in Eq.(\ref{a4}) and this minimum takes the value $1/15$. Hence we find that linear positivity is satisfied for all ranges of $t$ if
\beq
| \alpha | \le \frac {1} {15}
\eeq
There is therefore a non-trivial neighbourhood around the maximally mixed state, $\alpha = 0$, in which linear positivity is satisfied. Since generalized NSIT Eq.(\ref{con1}) is always satisfied by the quasi-probability, the combination of generalized NSIT plus linear positivity represents a clear improvement on the original NSIT condition Eq.(\ref{cond}) which can only be satisfied at $\alpha = 0$ but not in a neighbourhood of it. Hence the new framework described here is more robust than the original NSIT condition.

For values of $\alpha$ outside this range, linear positivity can still be easily satisfied, but for time ranges that are restricted. To characterize these ranges precisely would require a detailed solution of the inequalities Eqs.(\ref{a1})--(\ref{a4}), which may not be possible algebraically.
However, it is simpler and most relevant to focus on the range of values of $t$ close to $\omega t = \pi / 4 $ at which the maximum violation of the LG inequalities holds.

We easily find that, by inserting the value $\omega t = \pi / 4 $, Eqs.(\ref{a1}), (\ref{a2}) are satisfied at this point if
\beq
| \alpha | \le \sqrt{2} - 1
\eeq
and Eqs.(\ref{a3}), (\ref{a4}) hold if $ | \alpha | \le 1 $, which is true already. Hence linear positivity is easily satisfied at this single point.
Furthermore, since the quasi-probabilities depend continuously on $\alpha$, this means that for any fixed $\alpha$ with
$ |\alpha | < \sqrt{2}-1 $ there is a non-trivial time interval surrounding the point $ \omega t = \pi / 4 $, for which linear positivity is satisfied.

This model therefore confirms that there is a non-trivial class of mixed initial states for which linear positivity is satisfied and the LG inequalities are significantly violated, thus successfully implementing the protocol described earlier.

For pure initial states, for which linear positivity is not satisfied, one could consider a modified model in which the dynamics includes a decoherence mechanism described by a simple Lindblad evolution equation \cite{deco,Lin}.
This would create larger ranges in which linear positivity is satisfied (since it suppresses interference) but it would also lessen the violation of the LG inequalities (similar to the mechanism of disentanglement \cite{HaDo}). We have carried out calculations of a simple version of such a model. We find that the protocol still works in that it is possible to find ranges of time for which the LG inequalities are violated but both these ranges and the LG inequality violations are very small so this case may not be very relevant to experiment.

\section{Summary and Discussion}

Conventional approaches to understanding the LG inequalities are faced with either the theoretical and experimental difficulties of meeting the NSIT condition Eq.(\ref{cond}), or a more general version of NIM,
or the conceptual difficulties of interpreting them when this condition is not satisfied, in which case it is frequently asserted that a violation of the LG inequalities says more about the effect of measurement than about realism.
Although some promising experiments have been done \cite{Knee,Rob,Geo} which avoid these problems, it remains of interest to find alternative approaches.

Here we have proposed one such approach, which is to replace the usual two-time sequential measurement probabilities with measurable quasi-probabilities with very similar properties but which satisfy the analogue of the NSIT Eq.(\ref{cond}) exactly. We have shown that these quasi-probabilities have a number of properties
which can make them very useful in the study of macrorealism and the LG inequalities.

First, because of the way they are defined and measured, they describe the non-invaded part of the description of the system at two times. This is unlike the usual two-time measurement formula Eq.(\ref{2time}), in which $p(s_2)$ is affected by an earlier measurement, even if carried out using an ideal negative measurement. In particular, the fact that the quasi-probabilities satisfy generalized NSIT gives partial control over the degree to which invasive classical models can replicate the quantum results, but this to some degree also involves the specific measurement protocol used.

Second, the quasi-probabilities give a non-invasive indicator of MRps at two times. In particular, they are positive if and only if MRps holds. When positive, we obtain a natural parallel between the LG inequalities and Fine's theorem in the EPRB case. More generally, the quasi-probabilities together with the LG inequalities give a more refined account of macrorealism at two, three and four times and in particular highlight some situations in which the LG inequalities alone are insufficient.

We have shown in a simple model that parameter ranges are easily found for which linear positivity is satisfied but the LG inequalities are maximally violated.
Situations in which linear positivity is not satisfied with the LG inequalities either satisfied or not satisfied are also easily exhibited. This model also showed that the various interesting properties indicated by the quasi-probability can be satisfied in a robust way, unlike the original NSIT condition.

It will be of particular interest to check some of these ideas in experimental tests. This should be straightforward using simple modifications of existing experiments.

\section{Acknowledgements}

I am grateful to James Yearsley for useful conversations and also to Clive Emary for a careful reading of the manuscript and for useful suggestions. I also thank Oscar Dahlsten for a hosting a visit to Oxford to discuss these ideas, to Owen Maroney for useful discussions during that visit and to George Knee for discussions both during and after that visit and also for a critical reading of the manuscript.

This work was supported by in part by EPSRC grant No. EP/J008060/1.

\bibliography{apssamp}

\end{document}